# Graphene Growth by Metal Etching on Ru(0001)

E. Loginova[1], S. Maier[2], I. Stass[2,3], N. C. Bartelt[1], P. J. Feibelman[4], M. Salmeron[2], and K. F. McCarty[1i]

Sandia National Laboratories, [1]Livermore, CA and [4]Albuquerque, NM
[2]Materials Sciences Division, Lawrence Berkeley National Laboratory, Berkeley, CA
[3]Institut für Experimentalphysik, Freie Universität Berlin, Arnimallee 14, 14195 Berlin Germany

Low-energy electron microscopy (LEEM) reveals a new mode of graphene growth on Ru(0001) in which Ru atoms are etched from a step edge and injected under a growing graphene sheet. Based on density functional calculations, we propose a model wherein injected Ru atoms form metastable islands under the graphene. Scanning tunneling microscopy (STM) reveals that dislocation networks exist near step edges, consistent with some of the injected atoms being incorporated into the topmost Ru layer, thereby increasing its density.

## I. INTRODUCTION

The study of graphene on metal surfaces has a long history, culminating in widespread recent activity.[1] The work reported here focuses on the interaction of graphene with atomic steps and terraces on a precious metal surface. How graphene grows depends on the nature of this interaction. For example, on Ir(111)[2-5] and Ru(0001)[5-9], it mainly grows *over* atomic substrate steps forming a continuous carpet. On Ru(0001) this growth mode[7, 9] occurs by adding carbon to the free edge of the graphene sheet, e.g., point A in Fig. 1(b). (On Ir(111) the graphene also grows slowly up the staircase of substrate steps. [3, 5]) Carbon from a sea of carbon monomers (adatoms) attaches to this free edge. Our previous work suggests that the attachment occurs by an intermediate state involving several C atoms.[5, 7] Because of the high energy cost of forming these C clusters, carpet growth occurs only at high C supersaturation.

The difficulty of attaching individual carbon adatoms to the free graphene edge has another consequence – growth can occur by an additional complex mechanism, as illustrated in Fig. 1(b). Here the graphene sheet edge that abuts the substrate step edge (point B in Fig. 1(b)) advances by displacing atoms from the monatomic step of the substrate. The graphene sheet thus "eats into" the adjoining terrace. Evidence for graphene sheets embedded in terraces has been reported for graphene growth on Pt(111) [10-12], Pd(111) [13], and Ir(111) [3], but the mechanism by which the embedding occurs has not been determined.

In contrast to previous work, the results presented here show that carpet growth and etching processes occur on Ru(0001). Our focus is on the latter process. We have discovered that growth by Ru step etching results in the intercalation of the displaced Ru atoms under the graphene sheets. The intercalation alters the structure at the metal substrate/graphene interface and may change the properties of the film, e.g., its chemical reactivity.

## II. METHODS

We characterized graphene on Ru(0001) in two separate vacuum systems, employing low-energy electron microscopy (LEEM) in one and scanning tunneling microscopy (STM) in the other. LEEM was used to image the growth of graphene by segregating carbon from the Ru bulk or by decomposing ethylene, as previously described.[5, 7, 8] To evaluate whether the Ru steps had moved during graphene growth and removal, we typically compared the position of Ru substrate steps in the same surface region under three conditions: clean, after the local region had been completely overgrown by graphene and after its removal. This methodology accurately images the Ru step location by avoiding the local changes in image magnification that occur around the edges of graphene sheets. (These distortions result from the local variations in surface potential caused by the work-function difference between the bare and graphene-covered substrate.) Graphene was removed by either exposing to oxygen at elevated temperatures between 960 and 1040 K, or by thermally dissolving the graphene into the substrate.[8]

For STM studies, the Ru sample was cleaned by several annealing cycles between 800-1800 K in a partial oxygen atmosphere ($4\times10^{-8}$ Torr), followed by a short annealing period (20-30s) at 1700 K in ultra-high vacuum to desorb the remaining oxygen. Graphene was grown by segregation of carbon from the bulk. Under these conditions, graphene did not cover the entire surface but consisted of islands separated by several micrometers. Within the detection limit of Auger spectroscopy, no near-surface impurities were detected. Scanning tunneling microscopy was performed with a home-built instrument[14] with an RHK controller operated under ultra-high vacuum at 6 K.

## III. RESULTS

### A. Downhill growth of graphene over Ru steps

It has been reported[5, 7, 9] that on uniformly stepped Ru surfaces, graphene sheets preferentially nucleate at the lower edges of substrate steps and grow by extending across descending substrate steps. Figure 2 presents a sequence of LEEM images taken during graphene growth on a Ru morphology containing a large vacancy island, where the closed loop bounds the monatomic pit. Figure 2(a) and (b) show the Ru substrate before and after a graphene sheet has nucleated at the lower edge of a Ru step in the upper left corner. The graphene sheet then passes over the vacancy island in a continuous manner, apparently unhindered by the bounding step. After the region is completely overgrown (Fig. 2(d)), the metal step delimiting the vacancy island is still easily imaged through the graphene. Once covered by the graphene sheet, the vacancy island area and shape remained unchanged, within the resolution of our experiment.

### B. Graphene growth by Ru step etching

A strikingly different growth pattern occurs if graphene is nucleated *within* a vacancy island, as illustrated in Figure 3. Panel 3(a) is an image of a graphene-free surface with a vacancy island bounded by a monatomic Ru step. Figure 3(b) shows the nucleation of graphene inside that island, with the sheet abutting the Ru step edge. The graphene sheet quickly overgrows the vacancy island, as seen in Fig. 3(c). Thereafter, a graphene sheet that nucleated outside the field-of-view surrounds the sheet in the vacancy island, leaving the entire region covered with graphene. Figure 3(d) reveals that the area of the vacancy island has increased by about 40%. (To image the location of the Ru steps under the graphene more easily, the electron beam was tilted off axis for Fig. 3(d), a condition that creates contrast between Ru terraces separated by monatomic steps.[15]) The shape of the vacancy island has changed markedly, from a smooth oval to a hexagon, faceted along $<11\bar{2}0>$ Ru crystallographic directions. This shows that as the graphene grew within the island, Ru atoms were removed (etched) from the boundaries. In contrast, Ru was not removed when the vacancy island in Fig. 2 was overgrown by a graphene sheet draped over the Ru step. While the amount of Ru etching varied considerably with the local geometry of Ru steps and where graphene nucleated, the extent of step etching generally increased with temperature.

### C. Energetics of Ru etching

For step etching as shown in Fig. 3 to occur, the free-energy gain associated with graphene growth from C monomers (adatoms) must exceed the cost of displacing Ru atoms. The gain can be estimated from our previous work.[7] There, the formation energy needed to create a C monomer by removing a C atom from graphene on Ru(0001) was calculated (and measured) to be approximately 0.3 eV. A graphene sheet comprises about 2.4 C atoms per surface Ru atom. Thus, about 2.4 C atoms must be added to a graphene sheet edge growing into a Ru terrace for every Ru atom removed from the Ru step. Therefore, the energy required to displace a Ru atom must be below 2.4 × 0.3 eV = 0.7 eV. This value is considerably less than the formation energy of a Ru adatom on Ru(0001), which we calculate by density functional theory (DFT) to be about 1.3 eV. The energy of Ru adatoms on top of the graphene sheet should be even higher. (Our current estimate, based on new DFT calculations, is that the energy cost of incorporating a Ru atom from a Ru step into a 7-atom, regular hexagonal island on top of graphene/Ru(0001) is >2.5 eV.) This raises the question of where the ejected Ru atoms go when graphene grows into the Ru steps.

### D. Injection of Ru under graphene growing by Ru etching

One possibility is that Ru is displaced along the same Ru step that is being etched. This mechanism, pathway 3 in the schematic of Fig. 4, has been proposed to explain growth into terraces observed by STM on Pt[11, 12] and Ir.[11] The energy cost of moving the Ru atom along the step is small, and would allow graphene growth by etching to proceed by edge diffusion. However, in that case the area of a vacancy island being filled by graphene from within would not change, in contrast to what Fig. 3 clearly shows. On this basis, we infer that Ru atoms displaced from the steps move into the Ru terraces.

Obvious sinks for the liberated Ru atoms are upper terrace Ru steps (pathway 1 in Fig. 4). Indeed, we find that bare Ru steps near the steps bound to graphene collect displaced Ru atoms. Figure 5 provides an example. The two closed loops in image (a) bound vacancy islands. Some of the graphene growth in the field-of-view occurs by etching Ru steps, for instance, the vertical Ru step at the far right. After the region is overgrown with graphene, the two vacancy islands are noticeably smaller. (The area of the vacancy island on the left is reduced by a factor of six while the one on the right contracts by 40%. Presumably, the left-hand pit decreases more in area because it is exposed to the etching flux for the longest time.) From this evidence we infer that some of the Ru atoms are captured by adjacent, bare Ru steps.

Although emission of Ru atoms into the upper terrace is evident when there is a nearby bare Ru step (pathway 1 in Fig. 4), numerous Ru atoms also intercalate beneath the graphene sheet. The evidence is the rearrangement of Ru steps when graphene is subsequently removed – the intercalated Ru is released and reattaches to Ru step edges. This effect is demonstrated by following the step that bounds the central vacancy island in Fig. 6. Panel 6(a) shows the surface free of graphene at 1160 K. After cooling to allow C-atom segregation from the bulk, graphene nucleated and grew.[8] The image in Fig. 6(b) shows the surface at 965 K, about 7 hours later, when graphene completely covered the surface. The Ru step circumscribing the pit is readily imaged through the graphene layer.

Two striking changes occur in the pit perimeter. One is the faceting of the bounding step. The second is that the pit area increased by about 17% after graphene growth. Heating dissolves C back into the Ru (Fig. 6(c)) and the pit shrinks by about 18% (Fig. 6(d)), returning almost to its original size. No longer overgrown by graphene, the Ru steps are not faceted.

We interpret the recovery of the area of this vacancy island as evidence that as the C dissolved into the bulk numerous previously etched Ru atoms emerged from under it and reattached to the Ru step. In separate experiments, removing the carbon by reaction with $O_2$ also liberated intercalated Ru atoms.

The amount of intercalated Ru atoms released when graphene is removed strongly depends on the step configuration and increases with the substrate temperature. The largest amount of liberated Ru atoms was ~0.25 ML at 1200 K and ~0.03 ML near 1000 K for graphene grown inside isolated pits, consistent with our observation of more etching with increased temperature (see section 3.2).

To summarize our findings: graphene sheets that grow into Ru steps displace Ru atoms. The displaced Ru atoms can be captured by adjacent bare Ru steps, after diffusing over the upper Ru terrace (pathway 1 in Fig. 4), or they can be injected under the growing graphene (pathway 2 in Fig. 4). Our results suggest that the extent of injection can be as large as 10 to 20 % of a monolayer. [16] The observation that pits several microns across expand during etching growth and shrink with graphene removal suggests that the injected Ru atoms can diffuse far from their source/sink, the Ru step edge.

**E. Ru islands intercalated under graphene**

Ru adatoms are unlikely to remain isolated from one another under a graphene sheet. That would require the formation energy for an intercalated Ru atom to be < 0.7 eV (see section 3.3). Our DFT result is 1.2 eV (with the atoms in sites that continue the Ru-lattice (i.e., hcp) sites). Although slightly lower than the adatom formation energy in similar sites on clean Ru terraces (1.3 eV, section 3.3), this number is not nearly low enough to enable graphene formation by etching the Ru step edges.

A more plausible way to account for the observed etching, according to DFT, is formation of small Ru islands. On Ru(0001), graphene makes a moiré with the substrate and is periodically buckled.[6] As we describe next, Ru islands form with low energy under the moiré regions where the separation between the graphene sheet and the substrate is the largest.

There are two reasons that support this scenario. One is that with a 2.14 Å natural spacing of Ru(0001) layers, and the sum of the Ru metallic radius and the C covalent radius equal to 2.10 Å, buckling of the graphene layer to a calculated interlayer separation of 3.79 Å comes close to providing sufficient room for intercalated Ru atoms with no strain. The other reason is that the high point of the moiré occurs in its "atop region," where C atoms reside, approximately, above 3-fold hollows of the underlying Ru surface (and, surround atop sites – whence the nomenclature). This means that an under-graphene Ru atom in an hcp hollow will lie directly under a C-atom. Thus, these Ru atoms are well-positioned to interact with the C $2p_z$ orbital, just as the surface layer Ru atoms do with the C atoms directly above them in the low-lying regions of the moiré.

To test this idea we optimized islands of several sizes, producing a set of formation energies and chemical potentials. In each case, the atoms forming the Ru island were placed in hcp hollow sites, and the island, as a whole, was centered in the moiré atop region. An illustration of a moiré supercell containing a 19 Ru-adatom underlayer island is shown in Fig. 7. The numerical results of this study are reported in Table I.[17]

Even a small under-graphene island of seven Ru atoms has a low-enough Ru formation energy per atom (0.67 eV), compared to the 0.7 eV maximum needed for the growth of graphene, to form readily. Without graphene, the formation energy is 0.81 eV, higher than 0.7 eV, and thus such small Ru adatom islands would not be stable. Since the Ru chemical potential of all the under-graphene islands is less than 0.7 eV, they in principle would continue to grow once nucleated. We do not observe such growth with LEEM.

To understand why Ru adatom islands do not grow without limit, note in Table I that the chemical potential of the under-graphene islands starts to increase after they contain more than 19 atoms. Without the overlying graphene layer, the chemical potential would decay as the inverse of its radius (the usual Gibbs-Thomson effect) and the islands would Ostwald ripen to macroscopic size. Table I shows that is not the case for Ru under graphene. Thus, if nucleation barriers are small enough, ejected Ru might nucleate new small islands rather than attach to previously nucleated ones.

The amount of injected Ru atoms inferred from removing graphene from Ru pits (section 3.4), amounting to 10 - 20% of a monolayer, is roughly consistent with 19-atom islands underneath every moiré unit cell. Consider the example of Fig. 6, where a 17% area change was found. If the initial area of the pit was $A$, and new area of the pit is $A+dA$, then the percent coverage of new pit by intercalated Ru is $dA/(A+dA) = 0.17/1.17$

= 0.145. If the area of the moiré unit cell contains 11×11 or 121 Ru atoms, one 19 Ru underlayer island per cell amounts to a coverage of 19/121= 0.157 (see Table I).

**F. Experimental approaches to validate the existence of under-graphene Ru islands**

Why have intercalated islands not been detected so far with STM or X-Ray diffraction?[1, 18] We discuss two relevant issues, first, how prevalent intercalated Ru islands would be on a typical Ru surface, and second, how intercalated islands might affect STM images, particularly the corrugation in apparent height. We begin by noting that DFT predicts that the under-graphene islands are metastable compared to Ru attached to steps. (The island formation energies in Table I are positive.) Ultimately they will decay and their Ru content will end up at non-abutted step edges. Thus, this decay must be kinetically hindered to detect the under-graphene islands. In the LEEM experiments this hindrance came from two facts. First, we deliberately examined regions with very low step densities, with several microns between steps. Second, the largest amounts of etching and release, which are consistent with the model of under-graphene islands, occurred for graphene nucleated and grown within isolated monatomic pits (like Figs. 3 and 6). The step loop of the pit provides a type of closed system that prevents diffusion of under-graphene Ru to non-abutted Ru steps.

In contrast, the typical Ru surface consists of step arrays, not loops, with step spacings much smaller than the micron-sized terraces studied here by LEEM. We have not been successful in finding such sites in our STM studies, a reflection of the technique's difficulty in imaging large areas, relative to LEEM. Overall, under-graphene islands would only be expected at the rare locations where a graphene sheet abuts a Ru step separating relatively wide terraces. However, the current lack of direct observation of the under-graphene islands does not rule out their potential to serve as a low-energy intermediate state that allows graphene to etch Ru,

So, how should the under-graphene islands affect STM? In principle, atomically resolved STM could see the effect of the under-graphene islands – the islands convert atop sites of the non-intercalated moiré to hcp-like sites. Since resolving these subtle differences in STM is likely challenging, we next discuss the effect of the intercalated islands on STM height corrugation and propose an alternative characterization method.

Testing the hypothesis of intercalated Ru islands by examining the height corrugation of graphene in STM is difficult: DFT shows that the height difference between highest and lowest C atoms without intercalated Ru islands is 1.55 Å. With a 19-Ru island, the height difference is 2.18 Å. With a 37-Ru island, the corrugation is not very different, 2.22 Å. With a 7-Ru island, it is 2.07 Å. The lack of striking differences in corrugations between all these predicted moirés and the observed experimental moiré (with about 1 Å corrugation[1]) does not allow a definitive determination of the existence of underlayer islands, given the uncertainty of the relative importance of electronic and topology effects in determining the STM corrugation.

To confirm the existence of intercalated Ru experimentally one could deposit adatoms on top of the graphene, effectively conducting a titration. Recent studies of Ir deposition on a graphene moiré grown on Ir(111) shows that Ir adatoms cluster in regions of the moiré where every other C atom lies directly above an Ir atom of the underlying metal substrate.[19] The reason is that the graphene can then buckle locally, changing its bonding from $sp^2$ to $sp^3$ as it binds both to the adatoms above and the metal below.[20] This buckling does not occur in the "atop regions" of the moiré because all C atoms, there, reside above metal 3-fold hollows.

The situation is markedly different if there is a metal island under the graphene layer in its atop region (for the present discussion, on Ru(0001) rather than Ir(111)), as shown in the top view of Fig. 7. There, because the island atoms all sit in hcp 3-fold hollows, every-other C atom of the graphene layer now lies above an Ru atom, just as in the low-lying regions of the moiré. Thus, buckling of the graphene layer to bind adatoms is now energetically favorable, where it would not have been without the underlying island.

This discussion suggests that a "titration" can reveal the presence and size of under-graphene islands. With no such islands present, deposition of adatoms will leave the atop regions of the moiré bare, as found on Ir(111) by N'Diaye et al..[2] Where islands exist, adatom islands should form in atop regions.

In summary, DFT calculations suggest that under-graphene islands are a low-energy state that can accommodate large amounts of intercalated Ru. While the islands ultimately are metastable, they are an intermediate state that enables graphene to etch Ru steps by injecting Ru under graphene. Our LEEM observations of Ru etching and release are consistent with the proposal, but experimental confirmation awaits, perhaps using the approaches described above.

**G. STM evidence for another state of Ru injected under graphene**

The STM observations in Fig. 8 suggest another state of the intercalated Ru besides formation of islands. The graphene structures shown in the images are rotated with respect to the Ru substrate and the moiré periodicity is 1.4 ± 0.2 nm, smaller than the ~3 nm periodicity previously reported for the (12×12)C/(11×11)Ru[6, 7, 21-24] and the (25×25)[18] moiré superstructures. Distances in the STM images have been calibrated using the lattice spacing of bulk graphene (2.46 Å) as a reference. The angle between the moiré periodicity and the carbon lattice is $9 \pm 2°$ in Fig. 8(a, c, d). Small rotations of the graphene lattice relative to the substrate lattice can cause large changes in the moiré periodicity.[25] Based on the moiré equations derived by Nishijima et al.[26] we determined a rotation of the graphene relative to the ruthenium lattice of about 6° using a Ru-Ru distance of 2.71 Å.

Strikingly, we also observe a long-range triangular network with unusually large unit cells (side lengths 8.7 nm (Fig. 8(c)), 10.2 nm (Fig. 8(d)) and up to 12.8 nm in Fig. 8(b)). It is difficult to understand how this triangular network can result from placing graphene on a bulk-terminated Ru(0001) substrate. Instead, as we argue below, these triangles are consistent with a network of misfit dislocations that reconstructs the topmost Ru layer. The triangular networks were found only in graphene-covered regions of the substrate and in two separate growth (segregation) preparations. These triangular networks were only observed near Ru steps, and can be larger than 700 nm in extent in some areas or just local around a monatomic step, as shown in Fig. 8(b).[27] In this image, the terrace below the step has more triangles than the upper terrace. In addition, the Ru step is faceted along the close-packed $<11\bar{2}0>$ Ru directions, as in Fig. 3(d). The configuration around the Ru step could have resulted from a graphene sheet growing into the lower terrace followed by the growth of a separate sheet moving "downhill" on the upper Ru terrace.

The etched Ru atoms (pathway 2 in Fig. 4) can form the observed triangular networks by inserting themeselves into the topmost Ru layer, so that the first Ru layer has a slightly higher atomic density. The high-resolution STM images 8(c-d) reveal two types of triangular unit cells whose moiré patterns have different contrasts. We note that the imaging contrast of these structures was strongly voltage dependent. Adjacent cells are rotated by 180°. These observations suggest that one triangular cell type has the hcp stacking of the Ru while the other type has fcc stacking. Supporting this conclusion, the triangle orientation rotates 180° across a monatomic Ru step, as marked in the insert of Fig. 8(a). The lines in Fig. 8 are thus Shockley partial dislocations that separate areas where the topmost atomic layer has unfaulted hcp stacking from areas of faulted fcc stacking, similar to reconstructions reported on Pt(111)[28-30] during Pt homoepitaxy. The size of the triangles is then determined by how much Ru is taken up by the first Ru layer and the corresponding contraction of the lattice.

To explain the structure of the dislocations and to estimate how many Ru atoms are involved, we reproduce the triangle network with an atomic, two-dimensional Frenkel-Kontorova model of the topmost Ru layer.[4] We start with a moiré structure in which 41 Ru atoms are uniformly compressed to lie over 40 substrate atoms, yielding a unit cell size of 10.8 nm. Elastic relaxation in the top layer will concentrate the compression in regions away from the stable three-fold hollow sites of the substrate. To mimic this effect, we assume that nearest-neighbor Ru overlayer atoms interact through harmonic pairwise forces, while the substrate interaction is represented as a rigid sinusoidal two-dimensional potential.

Figure 9 shows a model configuration that schematically reproduces the structure measured with STM in Fig. 8. In particular, after relaxation, the simulation accurately reproduces the triangular-shaped dislocations in the white-shaded regions Fig. 9, as long as the energy difference between hcp and fcc regions is small enough that the areas of the two regions are similar. The dislocations in the Frenkel-Kontorova model lie along close-packed Ru directions, as also found in the STM images of Fig. 8.[31] While not included in the Frenkel-Kontorova model of Fig. 9, the graphene lattice on top of the dislocated Ru layer contributes the periodic corrugation observed in the STM images of Fig. 8.

While not completely proved, our model of the dislocation network in the topmost Ru layer well-describes the observed STM images. Since the model has extra Ru atoms in the topmost Ru layer, it provides evidence for the injection of Ru atoms underneath graphene. The validity of the model could be further evaluated using in-situ STM measurements. For example, whether the triangular networks formed as Ru steps were etched by graphene growth could be evaluated, as could their fate after graphene removal by oxygen exposure or dissolution into the bulk. In addition, the amount of Ru needed to create the proposed dislocation network is quite small − on the order of 0.1% of a ML. Our LEEM observations suggest that much larger amounts of Ru can be injected under the graphene where growth by the etching mode dominates over the carpet mode. Thus, the nature of the injected Ru atoms varies with their concentration. At low concentration, the injected Ru atoms may form dislocation networks in the topmost Ru layer. At high concentration, DFT calculations suggest that under-graphene Ru islands occur.

## IV. CONCLUSIONS AND SUMMARY

Our results show that graphene can grow on Ru(0001) by etching Ru steps. This growth mode is slower than graphene overgrowing descending Ru steps. On other substrates, the energetic barrier to attaching C atoms to a free graphene step edge may be even larger than for growth on Ru(0001).[7] Thus, etching could dominate growth on these substrates. Indeed, the etching mode also occurs during growth on Pt(111)[11, 12] and Pd(111).[13] Whether substrate atoms are also injected under graphene in these systems should be examined. The etching growth facets the Ru steps; therefore, graphene prefers particular bonding configurations at Ru steps. Step-etching also suggests that C bonds strongly to the Ru steps. Such tight bonding likely explains why graphene sheets do not grow up over the top of Ru steps, but etch the Ru step.

The etching growth mode has important consequences for the properties of the graphene film. The Ru atoms injected under the graphene sheet result in a more complex geometry than a flat, bulk-terminated substrate overlain by a buckled graphene sheet. When the concentration of injected Ru atoms is low, dislocation networks in the Ru layer under the graphene can occur, as seen by STM. At the higher concentration of injected Ru atoms observed by LEEM, islands of Ru atoms may exist under the graphene, as suggested by DFT calculations. Characterizing these structural changes is important for understanding the physical properties of graphene sheets. For example, the chemical reactivity of the graphene sheet could well be modified by intercalated islands. In fact, we suggest that titrating the graphene film with metal adatoms can distinguish local regions with and without intercalated Ru islands.

Lastly we remark on how unusual the etching mode is compared to typical film growth. Films commonly grow by atoms (molecules) attaching to the open site (a kink or a step) at the free edge of the film, because there the attachment barrier is zero or small.[32]

The ability of graphene to grow by displacing atoms on metal surfaces is another consequence of the exceptionally large barrier for direct adatom attachment.[7]

## ACKNOWLEDGEMENT

This work was supported by the Office of Basic Energy Sciences, Division of Materials Sciences and Engineering of the US DOE under Contracts No. DE-AC04-94AL85000 (SNL) and No. DE-AC02-05CH11231 (LBL).

[1] J. Wintterlin, and M. L. Bocquet, Surface Science **603**, 10 (2009).
[2] A. T. N'Diaye, S. Bleikamp, P. J. Feibelman, and T. Michely, Physical Review Letters **101**, 219904 (2008).
[3] J. Coraux, A. T. N'Diaye, M. Engler, C. Busse, D. Wall, N. Buckanie, F. J. M. Z. Heringdorf, R. van Gastei, B. Poelsema, and T. Michely, New Journal of Physics **11**, 023006 (2009).
[4] E. Loginova, S. Nie, N. C. Bartelt, K. Thürmer, and K. F. McCarty, Physical Review B **80**, 085430 (2009).
[5] E. Loginova, N. C. Bartelt, P. J. Feibelman, and K. F. McCarty, New Journal of Physics **11**, 063046 (2009).
[6] S. Marchini, S. Gunther, and J. Wintterlin, Physical Review B **76**, 075429 (2007).
[7] E. Loginova, N. C. Bartelt, P. J. Feibelman, and K. F. McCarty, New Journal of Physics **10**, 093026 (2008).
[8] K. F. McCarty, P. J. Feibelman, E. Loginova, and N. C. Bartelt, Carbon **47**, 1806 (2009).
[9] P. W. Sutter, J.-I. Flege, and E. A. Sutter, Nature Materials **7**, 406 (2008).
[10] O. Nakagoe, N. Takagi, and Y. Matsumoto, Surface Science **514**, 414 (2002).
[11] T. Fujita, W. Kobayashi, and C. Oshima, Surface and Interface Analysis **37**, 120 (2005).
[12] T. A. Land, T. Michely, R. J. Behm, J. C. Hemminger, and G. Comsa, Surface Science **264**, 261 (1992).
[13] S.-Y. Kwon, C. V. Ciobanu, V. Petrova, V. B. Shenoy, J. Bareo, V. Gambin, I. Petrov, and S. Kodambaka, Nano Letters, ASAP (DOI: 10.1021/nl902140j) (2009).
[14] T. K. Shimizu, A. Mugarza, J. I. Cerda, M. Heyde, Y. B. Qi, U. D. Schwarz, D. F. Ogletree, and M. Salmeron, Journal of Physical Chemistry C **112**, 7445 (2008).
[15] J. de la Figuera, J. M. Puerta, J. I. Cerda, F. El Gabaly, and K. F. McCarty, Surface Science **600**, L105 (2006).
[16] Larger relative amounts of step etching during graphene growth, such as in Fig. 3, result from the alternative pathway of Ru atoms diffusing onto upper Ru terraces, path 1 in Fig. 4.

[17] As the moiré cell changes size with small rotations of the graphene lattice relative to the substrate, as discussed in section 3.7, the size of the most-stable underlayer island may also change.

[18] D. Martoccia, P. R. Willmott, T. Brugger, M. Bjorck, S. Gunther, C. M. Schleputz, A. Cervellino, S. A. Pauli, B. D. Patterson, S. Marchini, J. Wintterlin, W. Moritz, and T. Greber, Physical Review Letters **101**, 126102 (2008).

[19] A. T. N'Diaye, S. Bleikamp, P. J. Feibelman, and T. Michely, Physical Review Letters **97**, 215501 (2006).

[20] P. J. Feibelman, Physical Review B **77**, 165419 (2008).

[21] Q. Dai, J. Hu, and M. Salmeron, Journal of Physical Chemistry B **101**, 1994 (1997).

[22] B. Wang, M.-L. Bocquet, S. Marchini, S. Gunther, and J. Wintterlin, Physical Chemistry Chemical Physics **10**, 3530 (2008).

[23] M. C. Wu, Q. Xu, and D. W. Goodman, Journal of Physical Chemistry **98**, 5104 (1994).

[24] H. Zhang, Q. Fu, Y. Cui, D. Tan, and X. Bao, Journal of Physical Chemistry C **113**, 8296 (2009).

[25] J. Coraux, A. T. N'Diaye, C. Busse, and T. Michely, Nano Letters **8**, 565 (2008).

[26] Y. Nishijima, and G. Oster, Journal of the Optical Society of America **54**, 1 (1964).

[27] The dislocation network in Fig. 8(b) is decorated by water adsorbed at 130K to enhance the contrast. The adsorption of water on graphene structures will be described in detail elsewhere.

[28] M. Bott, M. Hohage, T. Michely, and G. Comsa, Physical Review Letters **70**, 1489 (1993).

[29] M. Hohage, T. Michely, and G. Comsa, Surface Science **337**, 249 (1995).

[30] R. Pushpa, and S. Narasimhan, Physical Review B **67**, 205418 (2003).

[31] Atomically resolved images of the clean Ru substrate were used to determine in-plane directions.

[32] W. K. Burton, N. Cabrera, and F. C. Frank, Philosophical Transactions of the Royal Society of London Series a-Mathematical and Physical Sciences **243**, 299 (1951).

| Adatoms | $Z_{max}$(C) | $E_{form}$ | $\bar{\mu}$ | Def'n of $\bar{\mu}$ |
|---|---|---|---|---|
| 1 | 3.85 Å | 1.17 eV | 1.17 eV | $E_{form}(1)$ |
| 7 | 4.28 Å | 0.67 eV | 0.59 eV | $(7E_{form}(7) - E_{form}(1))/6$ |
| 19 | 4.35 Å | 0.42 eV | 0.28 eV | $(19E_{form}(19) - 7E_{form}(7))/12$ |
| 27 | 4.34 Å | 0.39 eV | 0.30 eV | $(27E_{form}(27) - 19E_{form}(19))/8$ |

| 37 | 4.35Å | 0.46 eV | 0.35 eV | $(37E_{form}(37) - 19E_{form}(19))/18$ |
|----|-------|---------|---------|---------------------------------------|
|    |       |         | 0.39 eV | $(37E_{form}(37) - 27E_{form}(27))/10$ |

**TABLE I.** Computed properties of representative under-graphene, Ru adatom islands, including the number Ru adatoms in the island, the height, $Z_{max}(C)$, of the highest C atom with the island present, relative to the average height of the Ru surface layer, the island formation energy, $E_{form}$, per Ru adatom, the average island chemical potential, $\bar{\mu}$, and the formula whereby each value of $\bar{\mu}$ was computed. Note that the 7, 19 and 37 atom islands are regular hexagons. The 27 atom island has sides alternating between 3 and 4 Ru atoms in length. These results were calculated in a Ru(0001)-11×11 supercell, containing 121 Ru atoms per layer. Thus, one N-adatom island per supercell amounts to a Ru adatom coverage of N/121.

**Figures**

FIG. 1. Schematic illustration of graphene growth on a precious-metal surface. (a) A graphene sheet (black) nucleated at a monatomic substrate step, on the lower terrace. In the carpet growth mode, the graphene sheet advances over descending substrate steps by adding carbon to the sheet's free edge, labeled A. (b) In the etching growth mode, the sheet grows in the opposite direction by etching substrate atoms, causing the substrate step that abuts the graphene sheet (point B) to retract.

FIG. 2. Sequence of LEEM images (3.5 μm field-of-view) taken during growth of graphene island, nucleated outside a pit, at 0 s (a), 180 s (b), 520 s (c), and 2370 s (d) on Ru(0001) at 970 K and $C_2H_4$ pressure $\sim 1 \times 10^{-8}$ Torr. Schematic representations of the height profiles along horizontal lines indicated by black and white arrows are shown below each LEEM image.

FIG. 3. Sequence of LEEM images (3.5 μm field-of-view) taken during growth of graphene island, nucleated inside a pit at 890 K, at 0 s (a), 50 s (b), 1010 s (c), and 1130 s (d) on Ru(0001) at 1050 K and $C_2H_4$ pressure $5 \times 10^{-9}$ Torr. LEEM image (d) obtained with the electron beam tilted from the surface normal to give contrast between adjacent Ru terraces. Schematic representations of the height profiles along horizontal lines indicated by black and white arrows are shown below each LEEM image.

FIG. 4. Schematic representation of possible paths for etched Ru atoms during "uphill" growth of graphene on Ru(0001): (1) to the upper terrace, (2) under graphene sheet, and (3) on the same terrace away from graphene.

FIG. 5. LEEM images (6 μm × 6 μm) of Ru(0001) during graphene growth at 1020 K at 0 s (a), 60 s (b), 100 s (c), 210 s (d), 280 s (e), and 550 s (f) at $C_2H_4$ pressure of $1\times 10^{-8}$

Torr. Schematic representations of the profiles along horizontal lines indicated by black and white arrows are shown below each LEEM image.

FIG. 6. Retraction and advance of Ru steps during graphene segregation and dissolution, respectively. Ru steps have been manually traced for clarity with the closed loop bounding a pit. a) Initial graphene-free surface at 1160 K. b) Same region at 965 K, nearly covered by a complete graphene layer. c) Graphene dissolution at 1195 K. Graphene islands image black. d) Surface free of graphene at 1195 K. Field-of-view is 9 μm.

FIG. 7. Schematic (top and side view) of a 19 Ru island under the "atop" region" of a graphene moiré on Ru(0001). White lines delimit the moiré supercell. Island Ru adatoms are colored magenta. First- and second-layer Ru atoms are colored cyan and red, respectively, enabling one to see that the hcp region of the moiré is to the right of the island, and the fcc region is to the left. C atoms are colored yellow to brown according to their computed heights above the Ru surface layer.

FIG. 8. STM images of graphene grown by segregation on Ru(0001) annealed to 1670 K. (a) Terrace completely covered with the triangular reconstruction. Insert demonstrates 180º rotation of the same type of triangular moiré cell on crossing a monatomic Ru step. (b) Triangular reconstruction around a Ru monatomic step, showing more reconstruction on the lower terrace. High-resolution images of the reconstruction with different triangular side lengths 8.7 nm (c) and 10.2 nm (d). Scan parameters are: (a) -150 mV and 11 pA; (b) -103 mV and 13 pA; (c) -151 mV and 17 pA; (d) -150 mV and 11 pA.

FIG. 9. Frenkel-Kontorova model of dislocations in the topmost layer of Ru(0001). Green- and red-shaded regions represent the surface Ru atoms in hcp and fcc binding sites, respectively, separated by Shockley partial dislocations.

---

[i] Corresponding author. mccarty@sandia.gov

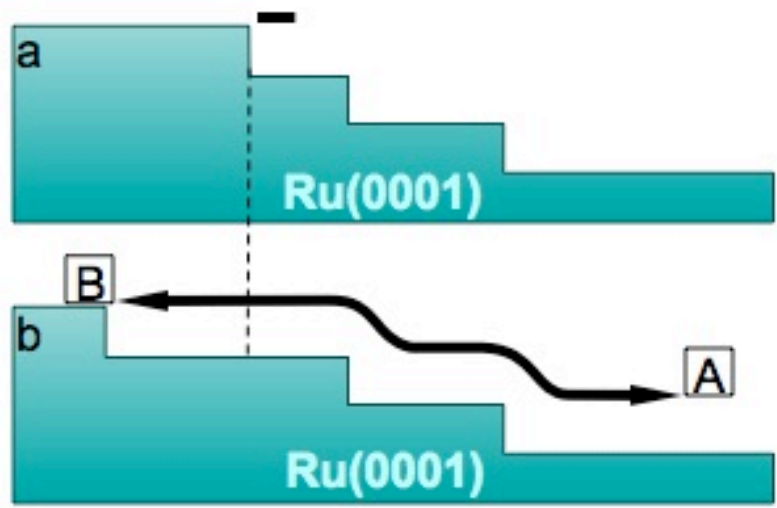

Figure 1.

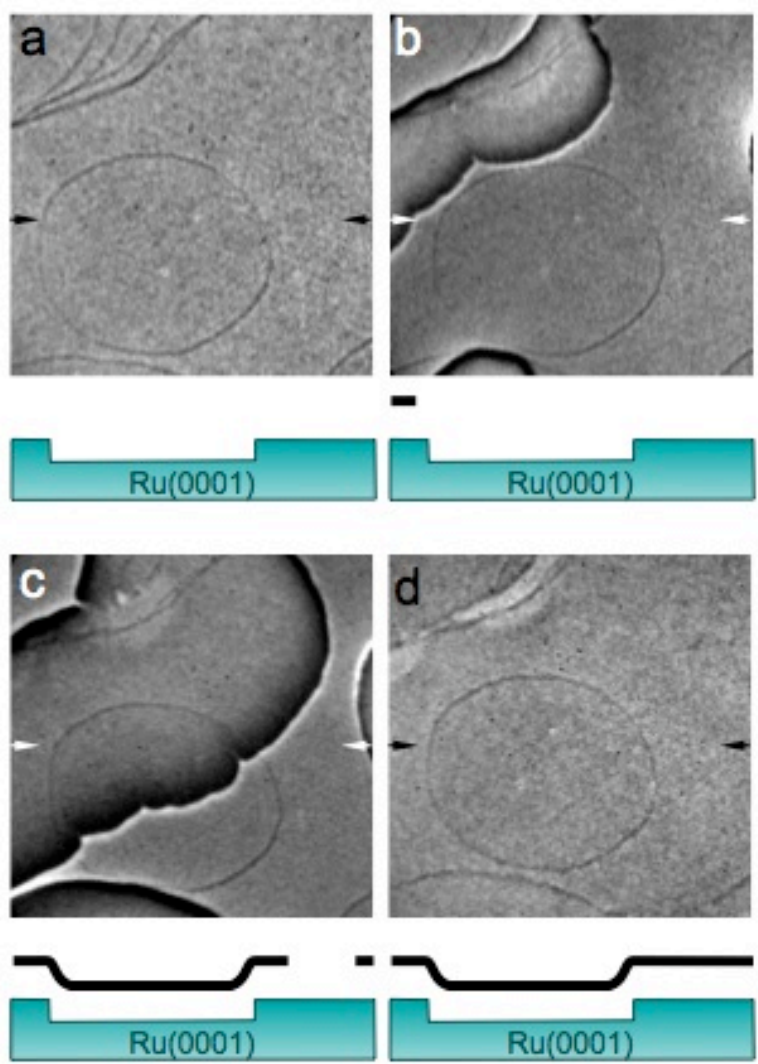

Figure 2.

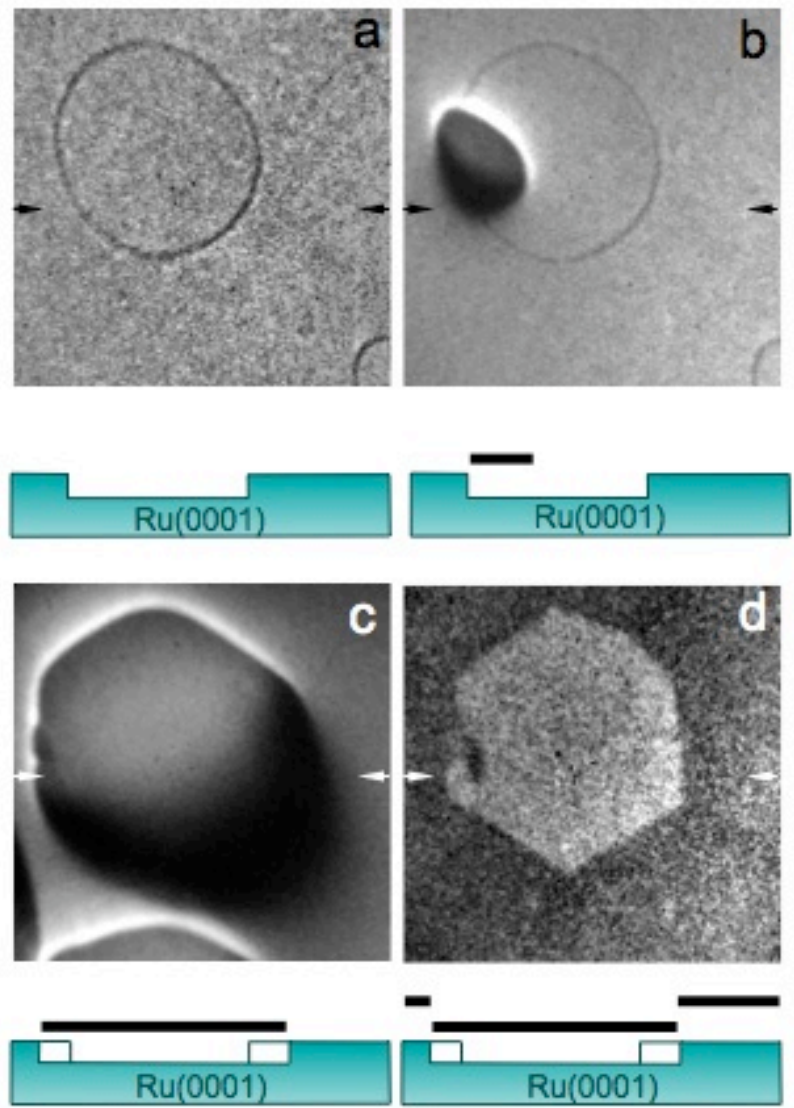

Figure 3.

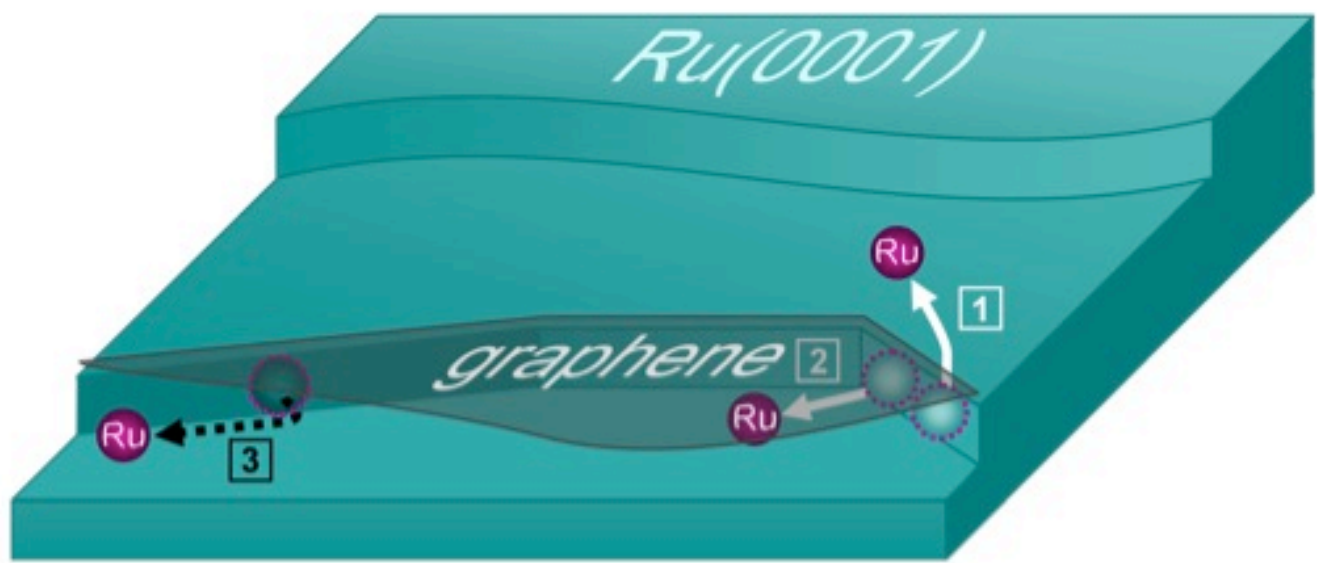

Figure 4.

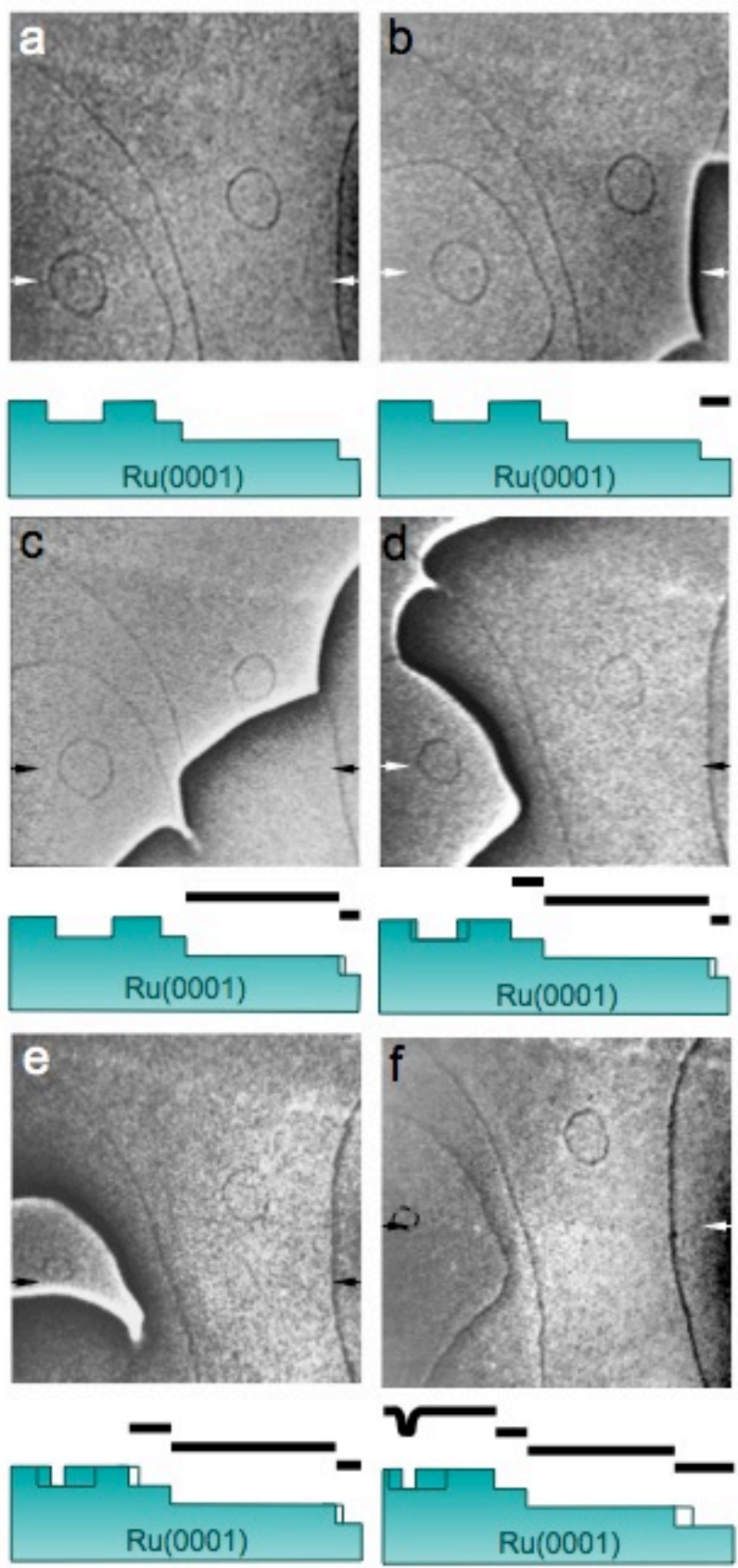

Figure 5.

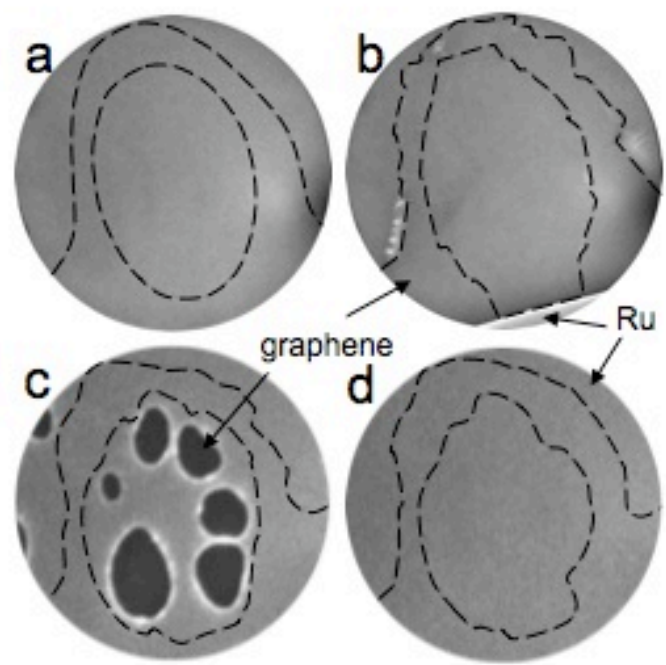

Figure 6.

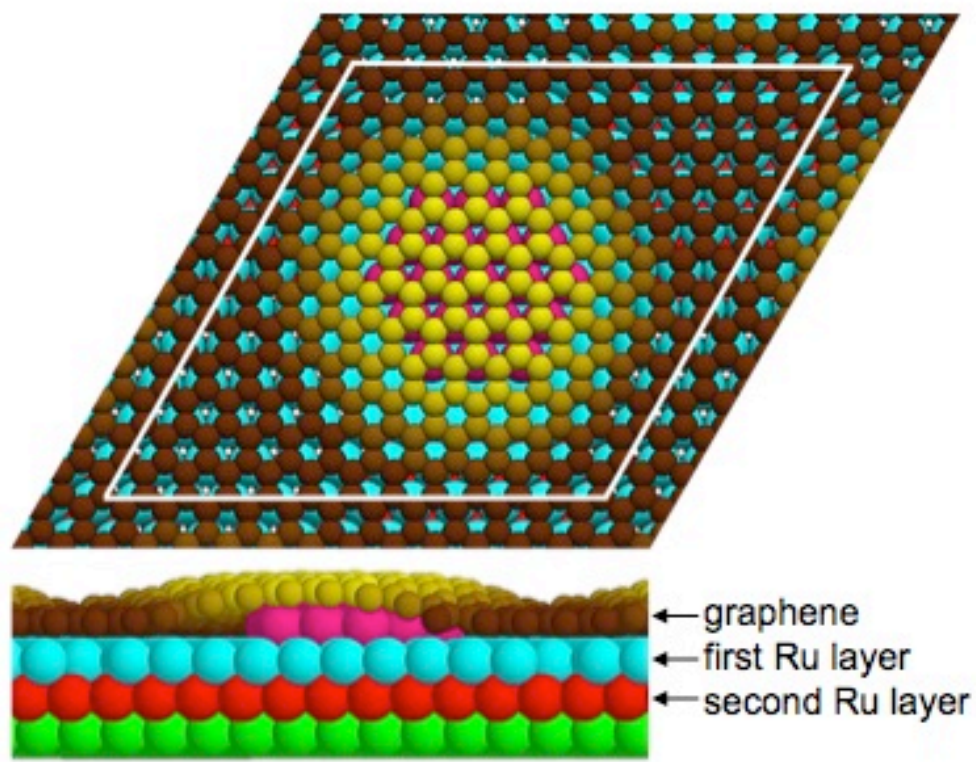

Figure 7.

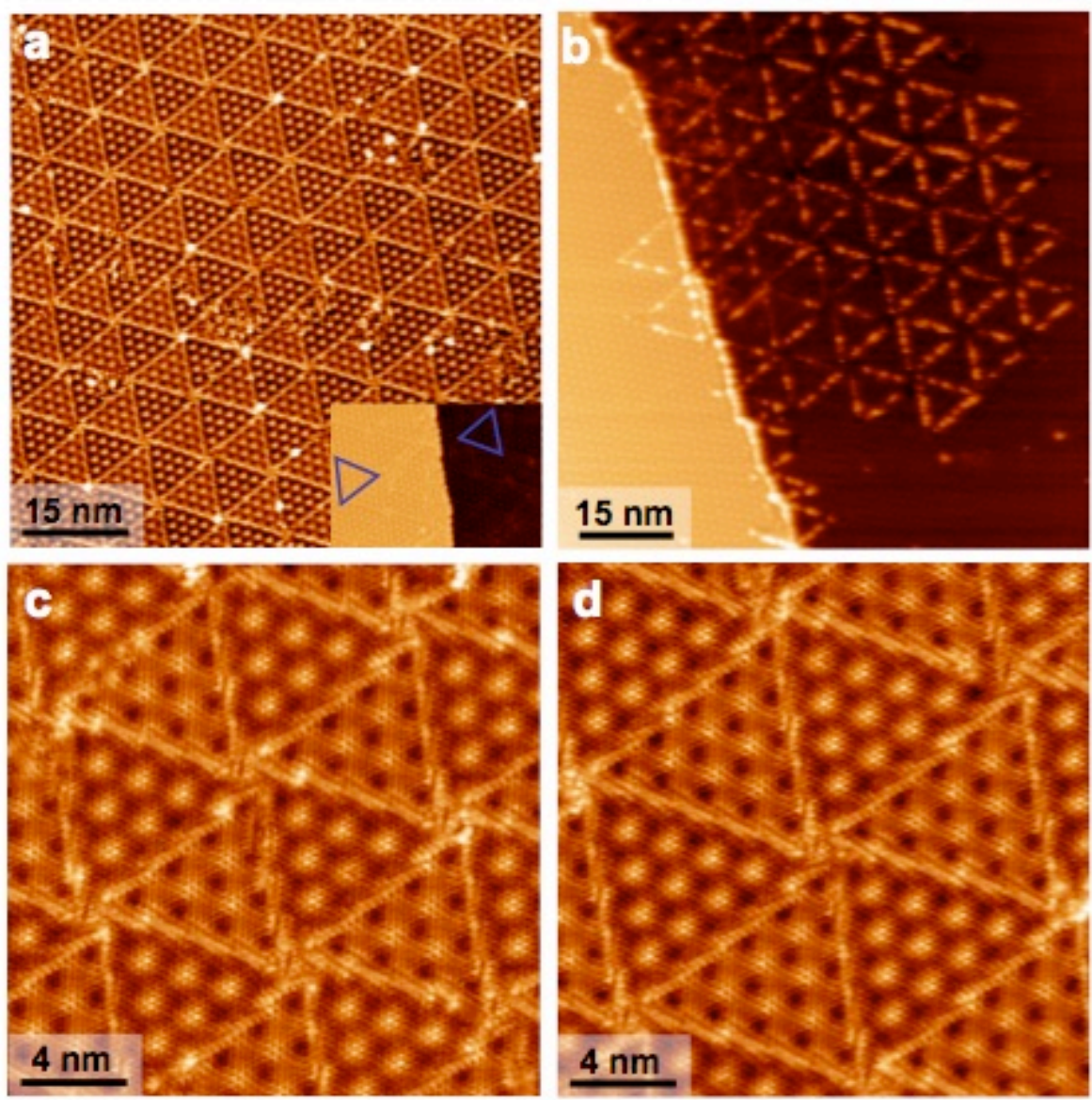

Figure 8.

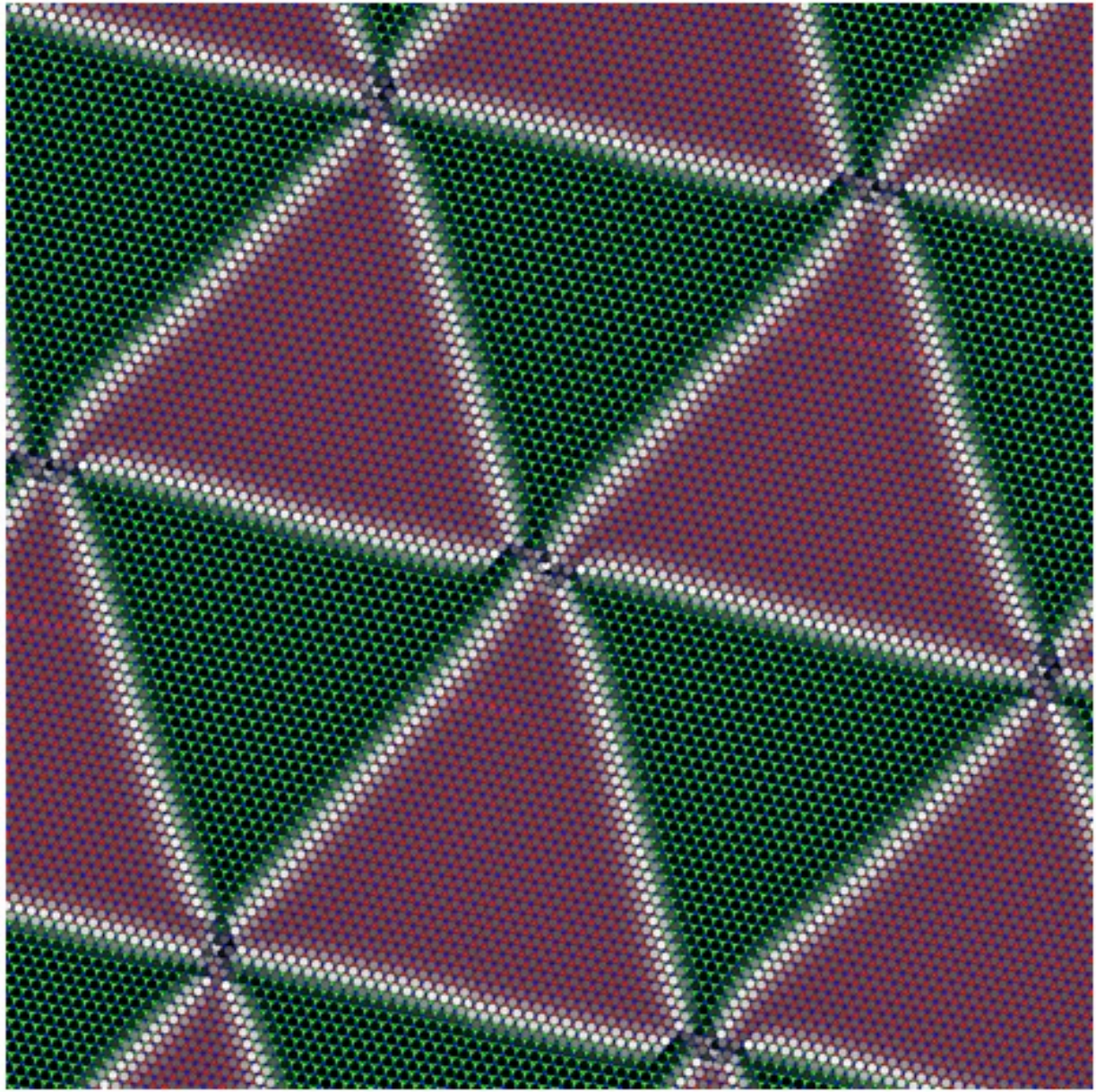

Figure 9.